\documentstyle[12pt]{article}
\newcommand{\extraspace}{\addtolength{\abovedisplayskip}{2mm}
                        \addtolength{\belowdisplayskip}{2mm}
                        \addtolength{\abovedisplayshortskip}{2mm}
                        \addtolength{\belowdisplayshortskip}{2mm}}
\newcommand{\be}{\begin{equation}\extraspace}

\newcommand{\ee}{\end{equation}}

\newcommand{\bea}{\begin{eqnarray}\extraspace}
\newcommand{\beastar}{\begin{eqnarray*}\extraspace}
\newcommand{\eea}{\end{eqnarray}}
\newcommand{\eeastar}{\end{eqnarray*}}

\newcommand{\nonu}{\nonumber \\[2mm]}

\newcommand{\strutje}{\rule[-1mm]{0mm}{4mm}}
\newcommand{\str}{\rule[-2.5mm]{0mm}{8mm}}
\newcommand{\stru}{\rule[0mm]{0mm}{6mm}}
\newcommand{\stro}{\rule[-2mm]{0mm}{5mm}}
\newcommand{\Str}{\rule[-3.5mm]{0mm}{9mm}}

\newcommand{\half}{{\textstyle \frac{1}{2}}}

\newcommand{\del}{\partial}
\newcommand{\eg}{{\it e.g.}}
\newcommand{\cW}{{\cal W}}

\newcommand{\ts}{\textstyle}

\begin{document}

\baselineskip=17pt

\hfill {PUPT-1442}

\hfill {hep-th/9401154}

\vskip 1.0cm
\begin{center}

{\LARGE Yangian Symmetry}\\
\vspace{4mm}
{\LARGE in}\\
\vspace{6mm}
{\LARGE Conformal Field Theory}\\

\vskip 2.0cm

{\large Kareljan Schoutens}

\vskip .7cm

\baselineskip=18pt

{\sl Joseph Henry Laboratories, Princeton University \\
     Princeton, NJ 08544, U.S.A.}

\vskip .7cm

{\bf Abstract}
\end{center}

\baselineskip=17pt

We show that the $SU(N)$, level-1 Wess-Zumino-Witten
conformal field theory provides a natural realization
of the Yangian $Y(sl_N)$ for $N\geq 3$. We also
construct a hamiltonian $H_2$ which commutes with the Yangian
generators and study its spectrum. Our results, which
generalize work by Haldane et al.\ \cite{hhtbp}, provide
the field theory extension of the algebraic structure
of the $SU(N)$ Haldane-Shastry spin chains with
$1/r^2$ exchange.

\vspace{.2cm}

\noindent
\baselineskip=17pt

\vfill
\noindent PUPT-1442

\noindent hep-th/9401154

\noindent January 1994

\newpage

\baselineskip=17pt

\paragraph{Introduction.}
In the early days of the Wess-Zumino-Witten (WZW)
conformal field theory, Polyakov and Wiegmann \cite{pol}
obtained a number of non-trivial results by applying
the Bethe Ansatz method to these theories. Since
that time, methods based on conformal invariance
have taken over, and have led to a detailed
understanding of these conformal field theories.
However, for certain applications, especially
those where a WZW model arises as an effective
description of a many body problem in condensed
matter physics, it would be desirable to have an
alternative approach to these theories, where by
`alternative' we mean an approach that does not
rely directly on the representation theory of
the Virasoro algebra or on the associated free field
realizations. Recently, Haldane et al.\ \cite{hhtbp}
proposed such a new approach to the $SU(2)$, level-1
WZW model. The underlying physical picture is that
of an ideal gas of `spinons', which obey fractional
statistics. The method of \cite{hhtbp} relies on the
{\it Yangian symmetry}\ in the $SU(2)$, level-1 WZW
conformal field theory, and we therefore expect that a
rather close connection with the Bethe Ansatz work of
\cite{pol} should exist.

We believe that it would be of definite interest to
investigate whether this new approach can be extended
to more general conformal field theories. As a first
step in this direction, we shall in this Letter
present the generalization of the field theory results
of \cite{hhtbp} to the $SU(N)$, level-1 WZW model for
$N\geq 3$.

\paragraph{Yangian symmetry.}  The Yangian
$Y({\bf g})$ associated to a Lie algebra ${\bf g}$
is a Hopf algebra that is neither commutative nor
cocommutative, and as such it can be viewed as a
non-trivial example of a quantum group \cite{drin}.
Its history goes back to the general formalism of the
Quantum Inverse Scattering Method (see \cite{vladimir}
for an introduction). Indeed, the object
$Y({\bf g})$ is directly related to certain rational
solutions of the Quantum Yang Baxter Equation, the
simplest of which was first obtained by Yang \cite{yang}.
In this Letter we focus on the case ${\bf g}=sl_N$.

We write the lowest generators of $Y(sl_N)$ as $Q_0^a$
and $Q_1^a$; higher generators can be obtained by taking
successive commutators with the generators $Q_1^a$.
The defining relations of the algebra $Y(sl_N)$ can be
written as follows \cite{drin}
\bea
{\rm (Y1)} && [Q_0^a,Q_0^b] = f^{abc} Q_0^c \ ,
\nonu
{\rm (Y2)} && [Q_0^a,Q_1^b] = f^{abc} Q_1^c \ ,
\nonu
{\rm (Y3)} && [Q_1^a,[Q_1^b,Q_0^c]] + ({\rm cyclic}\;\;{\rm  in}\ a,b,c)
     = A^{abc,def} \{ Q_0^d, Q_0^e, Q_0^f \} \ ,
\nonu
{\rm (Y4)} && [[Q_1^a,Q_1^b],[Q_0^c,Q_1^d] + [[Q_1^c,Q_1^d],[Q_0^a,Q_1^b]]
\nonu
&& \qquad
     = \left( A^{abp,qrs} f^{cdp} + A^{cdp,qrs} f^{abp} \right)
       \{ Q_0^q, Q_0^r, Q_1^s \} \ ,
\nonumber \eea
where $A^{abp,def} = {1 \over 4} f^{adp} f^{beq} f^{cfr} f^{pqr}$ and
the curly brackets denote a completely symmetrized product.
The $SU(N)$ structure constants $f^{abc}$ have been normalized
as
\be
   f^{abc} f^{dbc} = -2N \, \delta^{ad} \ .
\ee
The following comultiplications may be used to define
the action of the Yangian generators on a tensor product
of states
\bea
\label{copr}
&& \Delta_{\pm}(Q_0^a) =
  Q_0^a \otimes {\bf 1} + {\bf 1} \otimes Q_0^a \ ,
\nonu
&& \Delta_\pm(Q_1^a) =
  Q_1^a \otimes {\bf 1} + {\bf 1} \otimes Q_1^a
  \pm \half f^{abc} Q_0^b \otimes Q_0^c \ .
\eea
The `terrific' (dixit Drinfel'd \cite{drin}) right hand
sides of the relations (Y3) and (Y4) can be derived from
the homomorphism property of these comultiplications.
For ${\bf g}=sl_2$, the cubic relation (Y3) is superfluous
and for all other algebras (Y4) follows from (Y2) and (Y3).

\paragraph{Yangian Symmetry in Haldane-Shastry Spin Chains.}
In 1988, Haldane \cite{hal1} and Shastry \cite{shas}
proposed a class of integrable quantum spin chains that are
different from those that can be solved by means of the Bethe
Ansatz. A characteristic feature is that the spin-spin exchange
is not restricted to nearest neighbours. Instead, it has a
non-trivial dependence on distance, which, in the simplest
case, is of the form $1/r^2$. We shall here recall a few
aspects of the simplest of these models. We refer the reader
to \cite{hal2} for a recent account of the state of the art
in this field.

The hamiltonian $H_2$ of the $SU(N)$ Haldane-Shastry
chain with $1/r^2$ exchange acts on a Hilbert space that has
$N$ states for each site $i$, $i=1,2,\ldots,L$. It has
the form
\be
\label{Hxxx}
H_2 = \sum_{i \neq j} \left( {z_i z_j \over z_{ij} z_{ji}}
  \right) (P_{ij}-1) \ ,
\ee
where $P_{ij}$ is a permutation operator that exchanges the
states at sites $i$ and $j$, and $z_{ij}=z_i-z_j$. We choose
the complex parameters $\{z_j\}$ as $z_j=\omega^j$, with $\omega
= \exp(2\pi i/L)$, so that the exchange described by
(\ref{Hxxx}) is proportional to the inverse-square of the chord
distance between the sites. It was found in \cite{hhtbp} that
the hamiltonian $H_2$ commutes with the following operators
\be
\label{qq}
Q_0^a = \sum_i J_i^a \ , \qquad Q_1^a =
    {\ts {1 \over 4}} \, \sum_{i\neq j}
    {(z_i+z_j) \over z_{ij}} f^{abc} J_i^b J_j^c \ ,
\ee
where the $J^a_i$ are associated to the action
of $SU(N)$ on the $N$ basis states at site~$i$. Furthermore,
the operators $Q^a_0$, $Q_1^a$ satisfy the defining relations
(Y1) -- (Y4) of the Yangian $Y(sl_N)$. This remarkable result has
been understood to be at the basis of the integrability of these
spin chains. Indeed, it appears to be possible \cite{hhtbp}
to define mutually commuting integrals of motion $H_n$, $n\geq 3$,
which commute with the hamiltonian $H_2$ and with the Yangian
generators.

\paragraph{Extension to the $SU(2)$, level-1 WZW model.}
It was pointed out in \cite{hhtbp} that the algebraic structure
of the $SU(2)$ Haldane-Shastry spin chain has an interesting
extension to the case of the $SU(2)$, level-1 WZW conformal field
theory. The idea is that in the limit $L\rightarrow\infty$
the sites of the quantum chain become dense on the unit circle
and the summation $\sum_i$ may be replaced by a contour integral
$\oint {dz \over 2\pi i}$. The resulting operators act in the
chiral Hilbert space of the $SU(2)$, level-1 WZW conformal field
theory. This space is a direct sum of two irreducible
highest weight modules of the affine Kac-Moody algebra
$A_1^{(1)}$, which is generated by currents $J^a_m$ that satisfy
\be
\label{su2km}
[ J_m^a, J_n^b] = \delta^{ab} \, m \, \delta_{m+n}
  + f^{abc} J_{m+n}^c \ .
\ee
In the continuum limit, operator products are assumed
to be radially ordered, so that a double sum
$\sum_{i,j} f(z_i,z_j) J^a(z_i) J^b(z_j)$ (where $f$ is
some function of the $z$'s) is replaced by the double contour
integral $\oint {dz \over 2\pi i} \oint {dw \over 2\pi i}
f(z,w) J^a(z) J^b(w)$, where the contours ${\cal O}_z$,
${\cal O}_w$ encircle the origin and $|z|>|w|$.
This procedure leads to the following expressions for the
Yangian generators $Q_0^a$, $Q_1^a$ and $H_2$
\bea
\label{qqh}
&& Q_0^a = J_0^a \ , \qquad
   Q_1^a = \half \, f^{abc} \, \sum_{m>0} J^b_{-m} J^c_m \ ,
\nonu
&& \qquad H_2 = \sum_{m>0} m \, J^a_{-m} J^a_m \ .
\eea
These operators commute with the zero mode $L_0$
of the stress-energy tensor, which has the
usual Sugawara form $L_0 = {1 \over 2(N+1)} \sum_m
: J^a_{-m} J^a_m :$\ . It is important to notice that
$Q_1^a$ and $H_2$ are not zero modes of any conformal
fields in the theory.

In \cite{hhtbp} the following results for the operators
(\ref{qqh}) were conjectured%
\footnote{For the algebraic relations {\it (i)}\ and
{\it (ii)} to be valid the action of $Q^a_0$,
$Q_1^a$ and $H_2$ should be restricted to the
{\it irreducible} highest weight representations
of $A_1^{(1)}$ at level one.} \vspace{-2mm}
\begin{description}
\item[{\makebox[1cm][r]{\it (i)}}]
  $Q_0^a$ and $Q_1^a$ satisfy the generating
  relations of the Yangian $Y(sl_2)$,
\item[{\makebox[1cm][r]{\it (ii)}}]
  $Q_0^a$ and $Q_1^a$ commute%
\footnote{It is easily checked that the relation {\it (ii)}\
  is also satisfied for level $k=-2$; we have some indications
  that the irreducible highest weight representations at $k=-2$
  (or, for general $SU(N)$, $k=-N$) also carry a realization of
  the Yangian.}
  with $H_2$, and
\item[{\makebox[1cm][r]{\it (iii)}}]
  the spectrum (eigenvalues and eigenspaces) of the
  operator $H_2$ can be obtained by exploiting the
  representation theory of the algebra $Y(sl_2)$
  \cite{CP}.
  The eigenspaces can be characterized by half-infinite
  sequences of integer rapidities.
\end{description}

The authors of \cite{hhtbp} noticed that these field theory
results do not directly carry over to algebras different from
$SU(2)$. However, in the finite $SU(N)$ Haldane-Shastry spin
chains the Yangian $Y(sl_N)$ is realized as an exact symmetry,
and this clearly suggests that in $SU(N)$, level-1
conformal field theory a structure based on the Yangian $Y(sl_N)$
should exist. In the remainder of this Letter we shall show that
it is indeed possible to modify the expressions (\ref{qqh}) for
$SU(N)$ in such a way that properties similar to
{\it (i)}--{\it (iii)} [with $Y(sl_2)$ replaced by $Y(sl_N)$]
can be established. We shall actually prove that the modified
operators [given below in (\ref{qqhN}),(\ref{WW})] satisfy the properties
{\bf\it (i)} and {\bf\it (ii)}, and we shall check that, for the
first few levels, the spectrum of $H_2$ is consistent with the
`motif' prescription of \cite{hhtbp,hh}.

\vspace{8mm}
\paragraph{Generalization to $SU(N)$, $N \geq 3$.}
If one takes the expressions (\ref{qqh}) for $SU(N)$,
\hbox{$N\geq3$}, one may check that the cubic Serre relation
(Y3) of the Yangian and the relation
$[H_2,Q^a_1]=0$ are violated by terms that can be written
as zero-modes of conformal fields. This observation
suggests that in the case of $SU(N)$, $N\geq 3$, the
expressions (\ref{qqh}) should be corrected by extra
terms that are themselves zero-modes of conformal fields.
(There is clearly room for such terms, which could easily
appear if the limiting procedure that carries the
finite quantum chain over into the field theory would
be carried out with greater care.) One more piece of
intuition \cite{hal3} is that the extra term(s) in $H_2$
should be such that the degeneracy between conjugate
representations [say, the $10$ and the $\overline{10}$ of
$SU(3)$] is lifted, since the expected eigenvalues of
$H_2$ are different on these representations.

With this intuition, we arrived at the following Ansatz
for the Yangian generators and the hamiltonian
$H_2$
\bea
\label{qqhN}
&& Q_0^a = J_0^a \ , \qquad
   Q_1^a = \half \, f^{abc} \, \sum_{m>0}
       \left( J^b_{-m} J^c_m \right)
  - {\ts{\strutje N \over \strutje 2(N+2)}} \, W^a_0 \ ,
\nonu
&& \quad H_2 = \sum_{m>0} \left( m \, J^a_{-m} J^a_m \right)
  + {\ts{\strutje N \over \strutje (N+1)(N+2)}} \, W_0 \ .
\eea
These expressions contain the zero modes of the following
conformal fields (the brackets denote standard normal ordering,
see \eg\ \cite{bs} for conventions)
\be
\label{WW}
W^a(z) = \half d^{abc} (J^b J^c)(z) \ , \qquad
W(z) = {\ts{1 \over 6}} d^{abc} (J^a(J^bJ^c))(z) \ .
\ee
%
%
The 3-index $d$-symbol that occurs in these expressions
has been chosen to be completely symmetric and traceless
and has been normalized according to
\be
  d^{abc} d^{dbc} =
  {\ts{\strutje 2(N^2-4) \over \strutje N}} \, \delta^{ad} \ .
\ee
Notice that the extra term in $H_2$ is proportional to
the spin-3 generator of the $\cW_N$-algebra of the $SU(N)$,
level-1 WZW conformal field theory. It is to be expected
that the generators of spin $s \geq 4$ of these algebras
will appear as correction terms in the field theory expressions
for the higher conserved quantities $H_n$, $n\geq 3$.

We now claim that the operators defined in (\ref{qqhN}),
(\ref{WW}), when acting on irreducible highest weight
representations of $A_{N-1}^{(1)}$ at level one,
satisfy the properties {\bf\it (i)}--{\bf\it (iii)}
[with $Y(sl_2)$ replaced by $Y(sl_N)$] listed above,
and are thus the appropriate generalizations of the
$SU(2)$ operators given in (\ref{qqh}). The actual
verification of the algebraic properties {\bf\it (i)}
and {\bf\it (ii)} is rather involved, and we shall
only briefly explain these computations here. After
that we will discuss the spectrum of $H_2$.

Let us first focus on the defining relations of the Yangian
$Y(sl_N)$. For $N\geq 3$ the quartic Serre relation
(Y4) actually follows from the other relations and the
only non-trivial relation to be checked is the cubic relation
(Y3). If we write the left hand side of this relation as
\be
\label{lhsS3}
f^{bcd} [ Q_1^a , Q_1^d ] + ({\rm cyclic}\;\;{\rm  in}\ a,b,c)\ ,
\ee
we immediately see that terms in $[Q_1^a,Q_1^d]$ that are
of the form $f^{ade} X^e$ do not contribute. We may thus
compute modulo such terms. Using only the first terms in
$Q_1^a$ and $Q_1^d$, this gives the following contribution
to $[ Q_1^a , Q_1^d ]$
\be
\label{first}
  {\ts{1 \over 24}}
  \left( f^{abc}f^{ceg}f^{def} - f^{dbc}f^{ceg}f^{aef} \right)
  \left\{ (J^b(J^fJ^g))_0 - (J_0^g J_0^f J_0^b) \right\} \ .
\ee
The cross terms involving the first and the second terms in
$Q_1$ do not contribute; using the second term twice we
find a contribution
\bea
\label{second}
&& {\ts \left( {\strutje N \over \strutje 2(N+2)} \right)^2 }
   \left\{ \Str
   \left( d^{abc}d^{def}f^{ceg} - d^{dbc}d^{aef}f^{ceg} \right)
   \left( J^b(J^f J^g + J^g J^f) \right)_0 \right.
\nonu
&& \qquad \left. \Str
  + {\ts(N+2)} \, d^{abc}d^{dfc}
  \left( \del J^b J^f - \del J^f J^b \right)_0 \right\} \ .
\eea
It is easily checked that the term cubic in $J_0$ in (\ref{first})
precisely leads to the desired right hand side of (Y3). We should
thus prove that the sum of the remaining terms, when evaluated modulo
terms of the form $f^{ade} X^e$, vanishes. In order to proceed we
recall that on the Hilbert space of the $SU(N)$, level-1 WZW model
the following identity holds
\bea
\label{jajb}
(J^a J^b)(z) &=& \half f^{abc} \del J^c(z)
  + \, {\ts{\strutje 1 \over \strutje N^2-1}} \delta^{ab} (J^s J^s)(z)
\nonu
&& + \, {\ts{\strutje N \over \strutje (N^2-4)}} \, d^{abc} W^c(z)
   + \, {\ts {\strutje N^2 \over \strutje 288 (N-2)^2} } \,
   T^{ab,cd} V^{cd}(z) \ ,
\eea
where $W^c$ is as in (\ref{WW}) and $V^{cd}$ is defined as
\be
   V^{cd}(z) = \half \, T^{cd,ef} \, (J^eJ^f)(z) \ .
\ee
%
%
This identity implies that the symmetric product of the
currents $J^a$ and $J^b$ decomposes into three terms that transform
as a singlet (1), an adjoint representation $(N^2-1)$, and,
for $N \geq 4$, a third representation with Dynkin labels
$(010 \ldots 010)$ and dimension ${1 \over 4}(N-3)N^2(N+1)$.
The tensor $T^{ab,cd}$ that projects on this representation
is given by
\bea
\label{tensorT}
T^{ab,cd} &=& d^{abcd}
  + {\ts{\strutje 2N(N^2-9) \over \strutje (N^2-1)(N^2+1)}}
  \left[ 2 \delta^{ab}\delta^{cd}
  - {\ts(N^2 -1)}(\delta^{ac}\delta^{bd} + \delta^{ad}\delta^{bc}) \right]
\nonu
&& + {\ts{\strutje 2(N-3) \over \strutje N}}
   \left[ {\ts{\strutje 2N \over \strutje N^2-1}} \delta^{ab}\delta^{cd}
   + N (\delta^{ac}\delta^{bd} + \delta^{ad}\delta^{bc})
   + (f^{ace} f^{ebd} + f^{ade} f^{ebc}) \right] \ ,
\nonu
\eea
where we defined the completely symmetric, traceless 4-index
$d$-symbol as
\be
\label{4d}
d^{abcd} =
  d^{abe} d^{ecd} + d^{ace} d^{ebd} + d^{ade} d^{ebc}
  - {\ts {\strutje 4(N^2-4) \over \strutje N(N^2+1)}} \,
        (\delta^{ab} \delta^{cd} + \delta^{ac} \delta^{bd}
         + \delta^{ad} \delta^{bc}) \ .
\ee
By using the relation (\ref{jajb}) and a variety of contraction
identities among the tensors $f^{abc}$, $d^{abc}$ and $d^{abcd}$
one can work out the remaining terms in (\ref{first}) and
(\ref{second}). If one then uses that the following dimension-3
fields are both null,
\bea
\Phi^{ad}_3(z) &=& \left( f^{ahp}d^{pbd}-f^{dhp}d^{pba} \right)
   \left( J^b W^h \right)(z)
\nonu
   && + {\ts (N+2)} (\del J^d J^a - \del J^a J^d)(z)
\nonu
\Psi_3^{ad}(z) &=& {\ts{\strutje N^2 \over \strutje 288(N-2)^2}} \,
   \left( f^{abc} T^{cd,pq} - f^{dbc} T^{ca,pq} \right)
   \left( J^b V^{pq} \right)(z)
\nonu
   && - {\ts{\strutje N(N-3) \over \strutje N-2}}
      (\del J^d J^a - \del J^a J^d)(z) \ ,
\eea
one finds that indeed the remaining terms in (\ref{first}),
(\ref{second}) cancel. This then proves the validity of the cubic
Serre relation (Y3) and shows that the generators $Q_0^a$, $Q_1^a$
indeed generate the Yangian $Y(sl_N)$.

The fact that the operator $H_2$ defined in
(\ref{qqhN}), (\ref{WW}) commutes with the Yangian generators
can be established in a similar fashion. With a
certain amount of gymnastics, the commutator
$[Q^a_1,H_2]$ can be reduced to the zero mode of the
following dimension-4 conformal field
\bea
  \Xi_4^a(z) &=& {\ts{1 \over 6}} f^{abc} f^{cde}
        \left( \del J^b(J^d J^e) - J^b(\del J^d J^e) \right)(z)
\nonu
&& + \, {\ts{\strutje N^2 \over \strutje 4(N+2)^2}} \, d^{abc} \left[
      - {\ts{2 \over 3}} \left( W^b \del J^c +
                           \del J^b \, W^c \right)(z)
      + {\ts{1 \over 3}} \left( \del W^b \, J^c +
                           J^b \del W^c \right)(z) \right]
\nonu
&& + \, {\ts {1 \over 24} N(N-2)} \, \del^3 J^a(z) \ .
\eea
A lengthy but straightforward computation shows that this
conformal field is actually null, and this then establishes
that $H_2$ commutes with the generators of the Yangian.

\paragraph{The spectrum of $H_2$.}

The Hilbert space of the $SU(N)$, level-1 WZW conformal
field theory is a direct sum of $N$ irreducible highest
weight representations of the affine Kac-Moody algebra
$A_{N-1}^{(1)}$. Each of these `primary sectors' is
graded by the eigenvalue of $L_0$. The unitarity of these
representations guarantees that they are completely
reducible under the action of the Yangian, and that the
operator $H_2$ can be diagonalized. The Yangian generators
commute with $L_0$ so that the $L_0$ eigenspaces, which
are all finite dimensional, decompose into a finite sum of
(finite dimensional) representation of the algebra
$Y(sl_N)$. On each of these, $H_2$ (and similarly the
$H_n$, $n\geq 3$) takes a constant value.

The authors of \cite{hhtbp} proposed a scheme
for determining the eigenvalues and eigen\-spaces of $H_2$.
Adapted to the present case, it leads to the following
prescription. Each eigenspace is characterized by a
set $\{m_1,m_2,\ldots\}$ of distinct integer rapidities
$m_i>0$. These sets can be transformed into sequencses
of 0's and 1's, where a 0 (1) indicates the absence (presence)
of the integer that corresponds to the position in the sequence.
It is required that (i) at most $(N-1)$ consecutive 1's can
occur, and (ii) asymptotically the sequence stabilizes
on the pattern  11...11011...11011..., which alternates
between single 0's and bunches of $(N-1)$ 1's. There are
$N$ possible `phases' for the tail of a sequence, and these
correspond to the $N$ primary sectors of the $SU(N)$,
level-1 WZW model.

The rapidity sequences for states that are descendants w.r.t.
the affine algebra $A_{N-1}^{(1)}$ can thus be characterized
as finite rearrangements $\{ m_i^0\} \rightarrow \{m_i\}$
of the corresponding primary state sequence. The eigenvalues
of $L_0$ and $H_2$ are then determined as
\bea
&& L_0 = {\ts{\strutje 1 \over \strutje 2(N+1)}} \, C_2(R)
      + \sum_i \left[ m^0_i-m_i \right] \ ,
\nonu
&& H_2 = {\ts{\strutje N \over \strutje 6(N+1)(N+2)}} \, C_3(R)
      + \sum_i \left[ (m^0_i)^2 - (m_i)^2 \right] \ ,
\eea
where $C_2(R)$ and $C_3(R)$ are the eigenvalues of the Casimir
operators $J_0^aJ_0^a$ and \\
$d^{abc}J_0^aJ_0^bJ_0^c$ in the $SU(N)$ representation $R$
carried by the primary state $\{ m^0_i\}$.
It is expected that the eigenvalues of the $H_n$, $n\geq 3$
will be given by analogous expressions, which, for $n\leq (N-1)$,
will involve the eigenvalue of the order-$(n+1)$ Casimir operator of
$SU(N)$ in the representation $R$.

The paper \cite{hhtbp} also describes a procedure to break
up a rapidity sequence into a sequence of `motifs', which
can the be used to find an upper bound on the $SU(N)$
representation content of the corresponding Yangian multiplet.

We have studied the spectrum of the newly constructed
operator $H_2$ given in (\ref{qqhN}) and checked that its
eigenvalues and eigenspaces are consistent with the above
prescription. This once again confirms that the operators
(\ref{qqhN}) are the appropriate field theory extensions of
the finite chain expressions.
As an illustration, we present the results for the first few
levels of the three primary sectors for $SU(3)$ in Table~1,
and the results for the first four levels of the vacuum
representation for $SU(5)$ in Table~2.
For general $N\geq 4$, the eigenvalues of $H_2$ in the
vacuum module are $H_2=1$ for $L_0=1$, $H_2=4$, $2N+2$
for $L_0=2$, $H_2=5$, $9$, $2N+5$, $6N+3$ for $L_0=3$,
etc. Notice that the $SU(5)$ representation content $(24+75)$,
which occurs twice in Table 2, cannot be viewed as
a free product of $SU(5)$ representations (in the product
$10\times\overline{10}$ the singlet drops out). To our knowledge,
the systematics of the occurence of such `incomplete' multiplets,
which have been studied in \cite{hh}, has not been understood
completely.

\paragraph{Conclusions.}
In this Letter we proved that the operators $Q_0^a$
and $Q_1^a$ defined in (\ref{qqhN}), (\ref{WW}) generate the
Yangian $Y(sl_N)$, and that they commute with the
`hamiltonian' $H_2$. We also checked that the conjectures
of \cite{hhtbp} concerning the spectrum of $H_2$ are
consistent with an explicit evaluation of $H_2$ on
a number of low-lying states in the spectrum of
the $SU(N)$, level-1 WZW model.

An obvious question is how general the appearance of Yangian
symmetry in conformal field theory really is. Already at
the level of WZW models the situation is not clear. For
instance, if we stay at level $k=1$ but change the Lie algebra
to one that is different from $SU(N)$, we find that in general
the original Ansatz (\ref{qqh}) will not do. On the other hand,
modified expressions such as those in (\ref{qqhN}) cannot be
written, simply because algebras different from $SU(N)$ do not
have invariant 3-index $d$-symbols. The situation for higher
levels $k\geq2$ is equally unclear. One may expect that in
general rational conformal field theories a structure based
on a suitable generalization of Yangian symmetry may be
defined. Such a generalized Yangian would then be associated
to a general Chiral Algebra, rather than to a Lie algebra.
An obvious tool for investigating this issue would be the
Goddard-Kent-Olive coset construction. We leave these matters
for future investigations \cite{bs2}.

\paragraph{Acknowledgements.}
It is a pleasure to thank D.~Haldane for drawing my
attention to this problem and for illuminating discussions.
I also thank E.~Frenkel, Z.~Ha, V.~Korepin and in
particular P.~Bouwknegt for extensive discussions on this
topic. This research was supported by DOE grant
DE-AC02-76ER-03072.

\newpage

\begin{table}
\begin{center}
\begin{tabular}{||c|r|c|l|l||} \hline \str
$L_0$ & $H_2$ & $SU(3)$ irreps & $\{ m_i \}$ & `motifs' \\
\hline
\str
0 &  0   &  1     &  $\{1,2,4,5,7,8, \ldots\}$ & $(11)(11) \ldots$ \\
\stru
1 &  1   &  8     &  $\{2,4,5,7,8, \ldots\}$   & $(01)(11) \ldots$ \\
\stru
2 &  4   &  8     &  $\{1,4,5,7,8, \ldots\}$   & $(10)(11) \ldots$ \\
2 &  8   &  $1+8$ &  $\{2,3,5,7,8, \ldots\}$   & $()(11)(1)(11) \ldots$ \\
\stru
3 &  5   &  10    &  $\{4,5,7,8,,\ldots\}$     & $(00)(11) \ldots$ \\
3 &  11  &  $1+8_2
  +\overline{10}$ &  $\{1,3,5,7,8, \ldots\}$   & $(1)(1)(1)(11) \ldots$ \\
\stro
3 &  21  &  $1+8$ &  $\{2,3,5,6,8, \ldots\}$   & $()(11)(11)(1)(11) \ldots$\\
\hline
\str
${\ts{1\over3}}$ & $-{\ts{1 \over 9}}$ & 3 & $\{2,3,5,6, \ldots\}$
   & $()(11)(1) \ldots$ \\
\stru \stro
${\ts{4\over3}}$ & ${\ts{26 \over 9}}$ & $3+\overline{6}$
   & $\{1,3,5,6, \dots\}$ & $(1)(1)(11) \ldots$ \\
\hline
\str
${\ts{1\over3}}$ & ${\ts{1 \over 9}}$ & $\overline{3}$
  & $\{1,3,4,6,7 \ldots\}$ & $(1)(11)(1) \ldots$ \\
\stru \stro
${\ts{4\over3}}$ & ${\ts{10 \over 9}}$ & $6$
   & $\{3,4,6,7, \dots\}$ & $(0)(11) \ldots$ \\
\stro
${\ts{4\over3}}$ & ${\ts{46 \over 9}}$ & $\overline{3}$
   & $\{1,2,4,6,7 \dots\}$ & $(11)(1)(11) \ldots$ \\
\hline
\end{tabular}
\end{center}
\caption{Spectrum of $H_2$ on the lowest levels of the
  irrreducible highest weight representations of $A_2^{(1)}$
  in the $SU(3)$, level-1 WZW model.
  The eigenspaces of $H_2$ are irreducible representations
  of the Yangian $Y(sl_3)$. The `motifs', which have been
  constructed as in [1], encode information on
  the $SU(3)$ representation content of the Yangian
  multiplets, according to $(11)=1$, $(1)=\overline{3}$,
  $()=3$, $(0)=6$, $(01)=(10)=8$, $(00)=10$, etc.}
\end{table}

\newpage

\begin{table}
\begin{center}
\begin{tabular}{||c|r|c|l|l||} \hline  \str
$L_0$ & $H_2$ & $SU(5)$ irreps & $\{ m_i \}$ & `motifs' \\
\hline
\str
0 &  0   &  1        &  $\{1,2,3,4,6,7,8,9, \ldots\}$
                     &  $(1111)(1111) \ldots$ \\
\stru
1 &  1   &  24       &  $\{2,3,4,6,7,8,9, \ldots\}$
                     &  $(0111)(1111) \ldots$ \\
\stru
2 &  4   &  $24+75$  &  $\{1,3,4,6,7,8,9, \ldots\}$
                     &  $(1)(11)(1111) \ldots$ \\
2 &  12  &  $1+24$   &  $\{2,3,4,5,7,8,9, \ldots\}$
                     &  $()(1111)(111)(1111) \ldots$ \\
\stru
3 &  5   &  126      &  $\{3,4,6,7,8,9, \ldots\}$
                     &  $(0011)(1111) \ldots$\\
3 &  9   &  $24+75$  &  $\{1,2,4,6,7,8,9, \ldots\}$
                     &  $(11)(1)(1111) \ldots$ \\
3 &  15  &  $1+24_2+75 +\overline{126}$
                     &  $\{1,3,4,5,7,8,9, \ldots\}$
                     &  $(1)(111)(111)(1111) \ldots$ \\
\stro
3 &  33  &  $1+24$   &  $\{2,3,4,5,7,8,9,10,12, .\, .\}$
                     &  $()(1111)(1111)(111) \ldots$ \\
\hline
\end{tabular}
\end{center}
\caption{Spectrum of $H_2$ on the lowest levels of the
  vacuum representation of $A_4^{(1)}$ in the $SU(5)$,
  level-1 WZW model. The `motifs'
  are related to $SU(5)$ representations according to
  $(1111)=1$, $(111)=\overline{5}$, $(11)=\overline{10}$,
  $(1)=10$, $()=5$, $(0111)=24$, $(0011)=126$, etc. Note
  the occurence of the `incomplete' multiplets $(24+75)$,
  which arise as $10\times\overline{10}=1+24+75$, with
  the singlet dropping out.}
\end{table}
\end{document}